\documentclass[%
 reprint,
 amsmath,amssymb,
 aps,
]{revtex4-2}

\usepackage{graphicx}
\usepackage{dcolumn}
\usepackage{bm}
\usepackage{natbib}
\usepackage{hyperref}
\usepackage{xcolor}
\usepackage{graphicx} 
\usepackage{subfigure}

\begin{document}

\preprint{APS/123-QED}

\title{Heavy flavor particle production in p+p collisions at the CERN Large Hadron Collider energies}

\author{Tonmoy Sharma \textsuperscript{1}}
\email{tonmoy.sharma@cern.ch}
\author{Banajit Barman \textsuperscript{1}}
\email{banajit.barman@cern.ch}
\author{Buddhadeb Bhattacharjee \textsuperscript{1}}%
\email{buddhadeb.bhattacharjee@cern.ch}
\affiliation{%
\textsuperscript{1}Nuclear and Radiation Physics Research Laboratory, Department of Physics, Gauhati University, Guwahati, 781014, Assam, India
}%

\date{\today}

\begin{abstract}
The study of heavy-flavored baryon and meson production in proton–proton (pp) collisions provides crucial insight into the hadronization mechanisms of Quantum Chromodynamics (QCD). In this work an attempt has been made to describe the existing ALICE and LHCb results on the $p_\mathrm{T}$-differential production cross-section of $\Lambda_c^+$, $D^0$, $\Lambda_b^0$ and $B^0$ ($\bar{B}^0$), and $p_\mathrm{T}$-dependent $\Lambda_c^+$/$D^0$ and $\Lambda_b^0/B^0(\bar{B}^0)$ ratios, in light of the color-string junction inspired PYTHIA-8.314 model. A comparison of the results obtained with various datasets generated with different tuning parameters and the existing experimental results of ALICE and LHCb at $\sqrt{s}= $ 7 and 13 TeV underlines the relevance of string-junction dynamics in heavy-flavored particle production. The results of the present investigation reveal that while stopping the junction-antijunction reconfiguration in the non-perturbative PYTHIA-8.314 model successfully describes the considered ALICE experimental observables of the charm sector, on the other hand, allowing junction-antijunction reconfiguration, keeping other fragmentation parameters the same, better describes the LHCb experimental observables of both charm and bottom sectors.

\begin{description}
  \item[keywords]  heavy-flavor, baryon-to-meson ratio, PYTHIA, string junction, tuning parameters 
\end{description}
\end{abstract}

\maketitle

\section{\label{sec:introduction}Introduction}

String junctions are believed to be fundamental topological structures that can be produced and liberated in the nuclear collisions, and under appropriate orientation of the color field, the string junctions can migrate over a considerable rapidity space \cite{PhysRevC.111.014905}. With this argument, it was shown in the ref. \cite{PhysRevC.111.014905} that the MC model generated data of {\large P}YTHIA-8.310 \cite{10.21468/SciPostPhysCodeb.8, 10.21468/SciPostPhysCodeb.8-r8.3} string junction can successfully explain the experimental results of the $\bar{\Lambda}/ {\Lambda}$ ratio of both ALICE $(-0.8<y<0.8)$ and LHCb $(2.0<y<4.5)$ experiments for p+p collisions at $\sqrt{s} = $ 0.9 and 7.0 TeV. Further, it is well known that, unlike light-flavored quarks, the heavy-flavored quarks such as charm and bottom quarks are mainly produced in the initial hard scatterings of the collisions \cite{Norrbin2000-ij, Altmann2025-mj}. Such quarks, because of their large mass, cannot be produced during the hadronisation process. As a result, the total number of produced heavy flavor hadrons solely depends on the production rate of parent heavy quarks and remains unaffected by the hadronisation process \cite{Altmann2025-mj, Altmann2024-wm} or late-stage re-scatterings. However, the hadronization process determines the relative abundances of the various heavy-flavored hadron species and their kinematics \cite{Altmann2025-mj}. Given these properties of heavy-flavored hadrons, understanding their production mechanism becomes particularly intriguing. The ratio of a heavy-flavor baryon to a heavy-flavored meson serves as one of the key probes to gain insight into this process \cite{PhysRevC.109.054912, PhysRevLett.124.042301}. In contrast to light-flavored quarks, the heavy-flavor quarks, produced in the early stages of the collisions, traverse the dense QCD medium created in the nuclear collisions, thereby carrying the signatures of that medium \cite{RAPP_2010}. One notable signature is the observed enhancement of the heavy-flavor baryon-to-heavy-meson ratio in nuclear collisions compared to the vacuum-like electron-position collisions \cite{2018, PhysRevLett.127.202301, Acharya2019-vt, PhysRevLett.132.081901}. In ref. \cite{Altmann2024-wm}, it was shown that, in comparison to other studied models, the {\large P}YTHIA-8.311 string junction with the additional tuning parameters successfully described the ALICE results on the $p_\mathrm{T}$ dependence of the $\Lambda_c^+ / D^0$ ratio for p+p collisions at $\sqrt{s} = $ 13 TeV. However, no such result was found to be reported on the comparison of the $\Lambda_c^+ / D^0$ ratio in the LHCb acceptance. Further, the authors of ref. \cite{Altmann2024-wm} also studied the LHCb results on the $p_\mathrm{T}$ variation of the $\Lambda_b^0 / B^0$ ratio for p+p collisions at $\sqrt{s} = $ 13 TeV, but none of the studied tunes could give a proper description of the $\Lambda_b^0/B^0$ ratio. In this work, an attempt has been made with the various tunes of {\large P}YTHIA-8.314 p+p collisions data at $\sqrt{s}= $ 7 and 13 TeV, to discuss the experimentally measured ratios of heavy-flavored baryons to heavy mesons, particularly $\Lambda^+_c/D^0$ at both ALICE and LHCb acceptances, and $\Lambda^0_b/B^0$ at LHCb acceptance. It is worth mentioning here that no ALICE experimental results are available on the $\Lambda_b^0/B^0$ ratio. Guided by some genuine physics requirements, attempts have also been made to change the various parameters of the studied model to better describe the experimentally observed results of ALICE and LHCb.

\section{\label{sec:methodology}Methodology}

\subsection{\label{subsec:description_of_datasets}Description of the data set}
For completeness of the article, in this section, brief descriptions of the various tunes of {\large P}YTHIA-8.314  event generator and the parameters changed therein for the present investigation are narrated. {\large P}YTHIA-8.314 event generator is used to generate events of inelastic p+p collisions at $\sqrt{s} =$ 7 and 13 TeV, initially with {\large P}YTHIA Monash \cite{Skands2014-rk}(default), designated as {\large P}YTHIA\_MON, as a baseline. In this tune, the Lund String model for quark-diquark fragmentation \cite{Andersson1983-kb, SJOSTRAND1984469} is implemented along with the multi-parton-interaction (MPI) based color reconnection scheme \cite{PhysRevD.36.2019, Torbjorn_Sjostrand2004-vh}. A second dataset with the string junction model \cite{SJOSTRAND2003243, Christiansen2015-dt} of {\large P}YTHIA is generated with color Reconnection mode-1 (default) (via \verb|ColourReconnection:mode=1|) combined with the new beam remnant model (via \verb|BeamRemnants:remnantMode=1|) and is labeled as {\large P}YTHIA\_SJ. In this model, a new QCD-based color reconnection (QCDCR) scheme \cite{Christiansen2015-dt} is implemented, introducing a new mechanism of baryon-antibaryon pair production \cite{10.21468/SciPostPhysCodeb.8}. This mechanism involves the rearrangement of color flow in a multi-parton system, enabling three color-triplet quarks (or antiquarks) to be connected into a Y-shaped topological structure known as a string junction (anti-string junction) that favors 'shortest string length' \cite{Altmann2024-wm}. This is in addition to the default mechanism of baryon-antibaryon pair production via the string fragmentation to diquark-antidiquark pairs, as implemented in {\large P}YTHIA\_MON \cite{10.21468/SciPostPhysCodeb.8}. Physically, the string junctions represent the confinement field spanned between the three colored quarks (or antiquarks) in an overall color-neutral state \cite{Altmann2024-wm}. When freed from the attached quarks, this topological structure is believed to behave as a free-floating entity in the color field \cite{PhysRevC.111.014905} and act as the carrier of the baryon number \cite{PhysRevC.111.014905, 10.21468/SciPostPhysCodeb.8, Altmann2024-wm, Christiansen2015-dt}. The string junctions can be liberated from the incoming beam protons in the fragmentation regions from the remaining valence quarks of the beam remnants, making it possible that the charge and baryon numbers move independently \cite{PhysRevC.111.014905}. By default, this feature is disabled in the string junction model (color Reconnection mode-1) of {\large P}YTHIA, which leads to the production of diquarks from the beam remnants at extreme rapidities. However, {\large P}YTHIA allows us to switch on this feature by setting the flag \verb|BeamRemnants:beamJunction=on|, thereby liberating string-junctions and free quarks instead of diquarks from the beam remnants \cite{Lonnblad2023-xl}. The contribution of string-junctions from the beam remnants was previously found to be successful in describing the excess baryon over anti-baryon production in the LHCb acceptance, in particular \cite{PhysRevC.111.014905}. Hence, a third dataset with the same {\large P}YTHIA\_SJ, with the contribution of the \emph{liberated} string-junctions from the fragmentation regions,  is generated by enabling this flag. This dataset is designated as '{\large P}YTHIA String Junction with Remnants' Contribution' ({\large P}YTHIA\_SJ\_RC). The tuning parameters of all these datasets are summarized in the Table-\ref{table:tuningpar}.\\
Thirty million data sets were generated for each of the above settings for inelastic p+p collision events at $\sqrt{s} = $ 7 and 13 TeV.

\subsection{\label{subsec:calc_of_diff_prod_cs_and_intLum}Calculation of differential production cross-section and integrated luminosity}
With our generated {\large P}YTHIA datasets, the differential production cross-section of a particle as a function of transverse momentum $p_\mathrm{T}$ and rapidity $y$  was estimated by dividing the particle yield (as a function of $p_\mathrm{T}$ and $y$) by integrated luminosity $L_{int}$, that is,
\begin{equation}
    \frac{d^2\sigma}{dp_Tdy} = \frac{1}{L_{int}} \times  \frac{d^2 N}{dp_Tdy}
\end{equation}
where, $L_{int}$ is defined as,
\begin{equation}
    L_{int} = \frac{\text{Total number of events}}{\text{Cross-section$_{(inelastic)}$}} = \frac{N_{ev}}{\sigma_{INEL}}
\end{equation}
Here, $\sigma_{INEL}$ corresponds to the cross-section of inelastic p+p collisions, whose value is obtained from the {\large P}YTHIA-8.314 event generator for the studied two different center-of-mass energies $\sqrt{s} = $ 7 and 13 TeV, and is shown in Table-\ref{table:inelsigma}. The calculated values of integrated luminosities are also shown in the Table-\ref{table:inelsigma}.

\begin{table*}
\caption{\label{table:inelsigma}Values of inelastic cross-section and integrated luminosity for the generated {\large P}YTHIA8.314 datasets}
\begin{ruledtabular}
\begin{tabular}{c c c c}
Collision system& No. of events& $\sigma_{INEL}$ $\left(mb\right)$& $L_{int}$ $\left(\mu b\right)^{-1}$ \\
\hline
$pp_{\textbf{INEL}}$ $\sqrt{s} = $ 7 TeV& $30\times10^6$& 71.07& 422.119 \\
$pp_{\textbf{INEL}}$ $\sqrt{s} = $ 13 TeV& $30\times10^6$& 78.07& 384.270 \\
\end{tabular}
\end{ruledtabular}
\end{table*}

\subsection{\label{subsec:classfn_prompt_nonprompt_charm}Classification of prompt and non-prompt charmed hadrons}
Charm hadrons can be classified according to their production mechanism as prompt or non-prompt. Prompt charmed hadrons are those that are produced directly in collision through the hadronisation process or through the decays of excited open charmed and charmonium states that occur within the strong interaction time-scale (~$10^{-23}$ s). On the other hand, non-prompt charm hadrons are those that are produced through the weak decays of the bottom hadrons  \cite{Acharya2024-wf, mallick2025charmhadronreconstructionbodydecay} and other heavier charm hadrons.
In this study, the prompt and non-prompt $\Lambda_c^+$ and $D^0$ are classified in the {\large P}YTHIA-generated datasets by identifying their mother particles. The non-prompt  $\Lambda_c^+$ and $D^0$, whose mother particles have PDG particle identification code (PID) associated with the bottom and other heavier charmed hadrons, such as PID 5122, 5112, 5212, 5222, 5132, 5232, 5332, 5142, 5242, 5342, 5442, 5512, 5522, 5532, 5542, 5554, 4222, 4212, 4112, 4232, 4132, 4332, 4412, 4422, 4432, 4444, 511, 521, 531, 541, 411 and 431, are filtered from the dataset and are not considered for further analysis  \cite{ParticleDataGroup:2024cfk}.

\section{\label{sec:results_and_discussions}Results and discussions}
To gain insight into the production mechanism of heavy-flavored hadrons, {\large P}YTHIA-generated datasets of different tunes were compared with the various published experimental results of the ALICE \cite{2018, PhysRevC.94.054908, PhysRevLett.128.012001} and the LHCb \cite{PhysRevLett.132.081901, 20131, Aaij_2016} experiments. In sub-sections (\ref{subsec:results_charm} and \ref{subsec:results_bottom}), we report the results of our studies on the charmed sector and the bottom sector, respectively. The last sub-section (\ref{subsec:results_robustness_check}) focuses on the effect and validation of the models, in the light of results of our earlier works on $\Lambda$ baryons, reported in ref.\cite{PhysRevC.111.014905}.

\subsection{\label{subsec:results_charm}Comparison of \textbf{$\Lambda_c^+$} and \textbf{$D^0$} experimental results with model estimations}
This section focuses on the differential production cross-section of prompt $\Lambda_c^+$ and $D^0$, as measured by ALICE \cite{2018, PhysRevC.94.054908} ($|y|<0.5$) and LHCb \cite{20131} ($2.0 < y < 4.5$) collaborations in p+p collisions at $\sqrt{s} = $ 7 TeV. It is noteworthy that while the results of the ALICE collaboration on the production of $\Lambda_c^+$ and $D^0$ are the average of the particles and their charge conjugates \cite{2018, PhysRevC.94.054908}, for the LHCb experimental data points, it is the sum of the two \cite{20131}. The LHCb Collaboration has reported measurements on the $p_\mathrm{T}$-differential production cross section of $D^0$ over five consecutive rapidity intervals ($2.0<y<2.5$, $2.5<y<3.0$, $3.0<y<3.5$, $3.5<y<4.0$, $4.0<y<4.5$). Therefore, to obtain the $p_\mathrm{T}$-differential production cross-section of $D^0$ across the rapidity interval of $2.0 < y < 4.5$, the differential production cross-section values presented in Table 7 of ref.\cite {20131} were summed across the five rapidity intervals for each specified $p_\mathrm{T}$ bin. The associated statistical and systematic uncertainties were summed using the quadrature method. Furthermore, we have multiplied the LHCb measurement on $\Lambda_c^+$ production by a correction factor \cite{2018}, which takes into account the most recent value of the branching ratio (BR) for the $\Lambda_c^+ \longrightarrow pK^-\pi^+$ decay \cite{ParticleDataGroup:2024cfk}. Given that the BR term appears in the denominator of the differential production cross-section expression, the correction factor is obtained as follows:
\begin{equation}
   \frac{(d \, \sigma / d \, p_T)_{new}}{(d \, \sigma / d \, p_T)_{old}}= \frac{\text{BR}(\Lambda_c^+ \longrightarrow pK^-\pi^+)_{old}}{\text{BR}(\Lambda_c^+ \longrightarrow pK^-\pi^+)_{new}} = \frac{5.0}{6.24} = 0.80
\end{equation}
This correction is specifically limited to LHCb results on $\Lambda_c^+$ production. For all other experimental results \cite{2018, PhysRevC.94.054908, PhysRevLett.128.012001, PhysRevLett.132.081901, 20131, Aaij_2016} used in the present study, the deviation of the previously reported BR from the most recent measurement documented in ref. \cite{ParticleDataGroup:2024cfk} remains within 2 percent. This variation falls within the established bounds of the uncertainties associated with the experimental data points, making further correction irrelevant in these cases. It is to be noted here that the uncertainty linked to the recent BR value has not been considered in the correction of LHCb results on $\Lambda_c^+$ production. 

In the context of the aforementioned collision system, the ALICE measurement on prompt $\Lambda_c^+$ production was compared in ref. \cite{2018} with POWHEG+{\large P}YTHIA6 model and GM-VFNS scheme. These models showed a significant underestimation of the experimental data. On the other hand, the theoretical predictions from the GM-VFNS and FONLL schemes have been more or less successful in describing both the ALICE and LHCb measurements for prompt $D^0$ production \cite{PhysRevC.94.054908, 20131}. Additionally, regarding the LHCb measurements of prompt $\Lambda_c^+$ production, the predictions from the GM-VFNS scheme were found to be broadly consistent with the experimental data, considering the associated experimental and theoretical uncertainties \cite{20131}. However, due to the considerable uncertainties tied to these models and schemes, a conclusive remark on the agreement with the experimental data could not be made. The present investigation aims to gain a deeper understanding of the underlying production mechanisms of $\Lambda_c^+$ and $D^0$ hadrons by utilizing the more recent version of {\large P}YTHIA-8.3, i.e. {\large P}YTHIA-8.314 MC event generator, with reduced associated uncertainties. 

\begin{figure*}
  \subfigure[]{
    \includegraphics[width=0.45\linewidth]{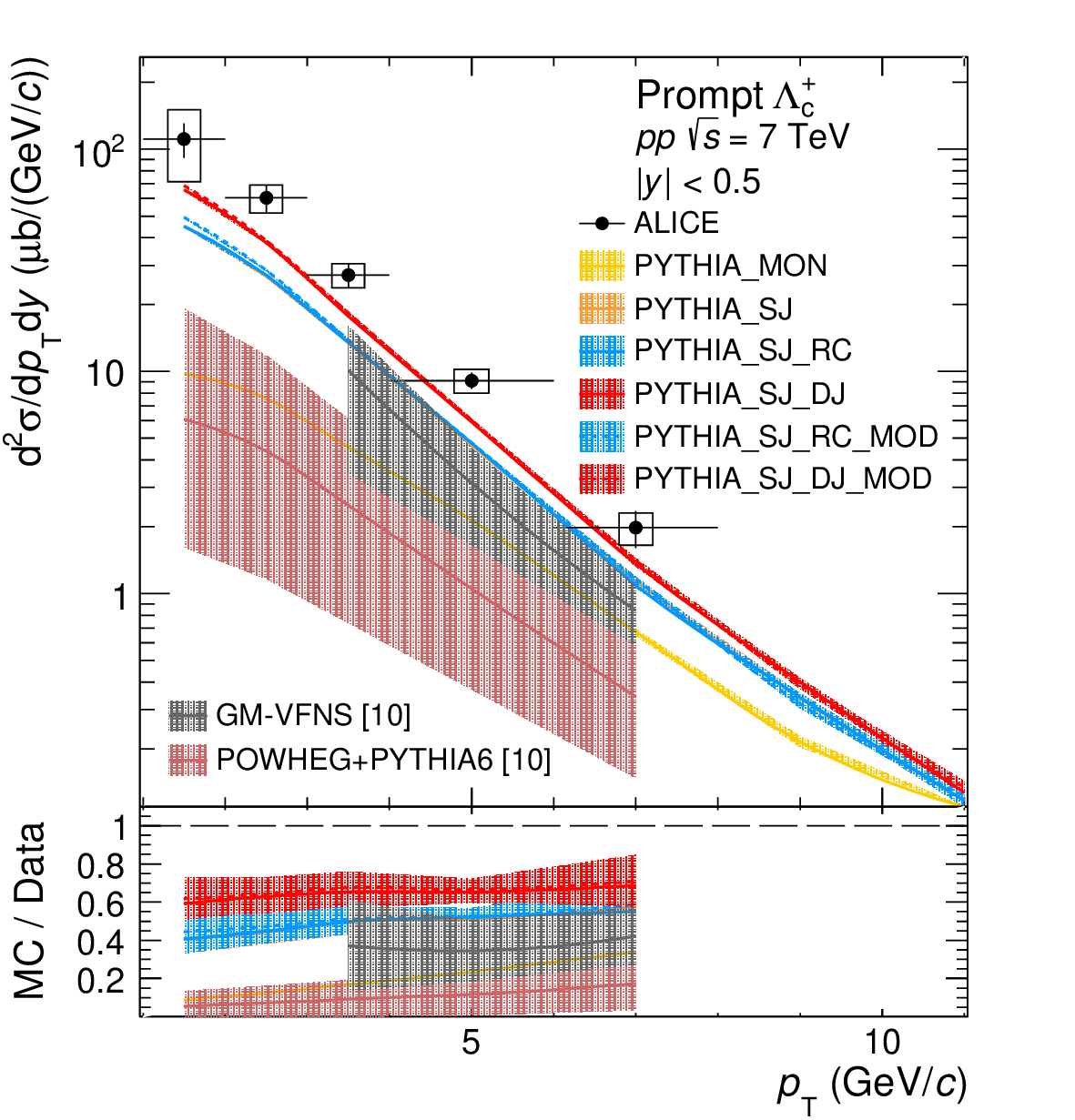}
    \label{fig:lc_pt_alice}
    
  }
  \subfigure[]{
    \includegraphics[width=0.45\linewidth]{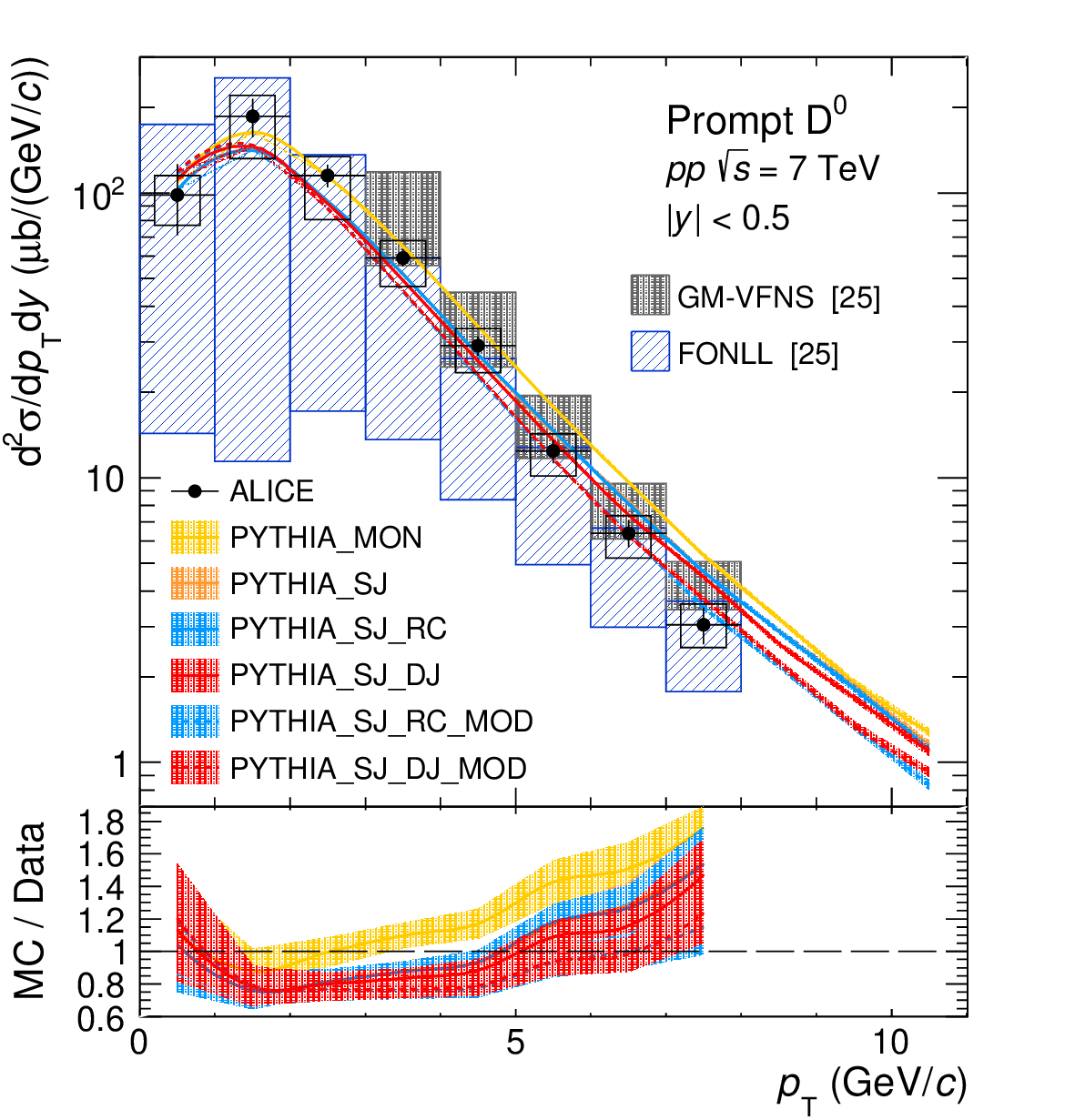}
    \label{fig:d0_pt_alice}
  } 
  \hfill
  \subfigure[]{
    \includegraphics[width=0.45\linewidth]{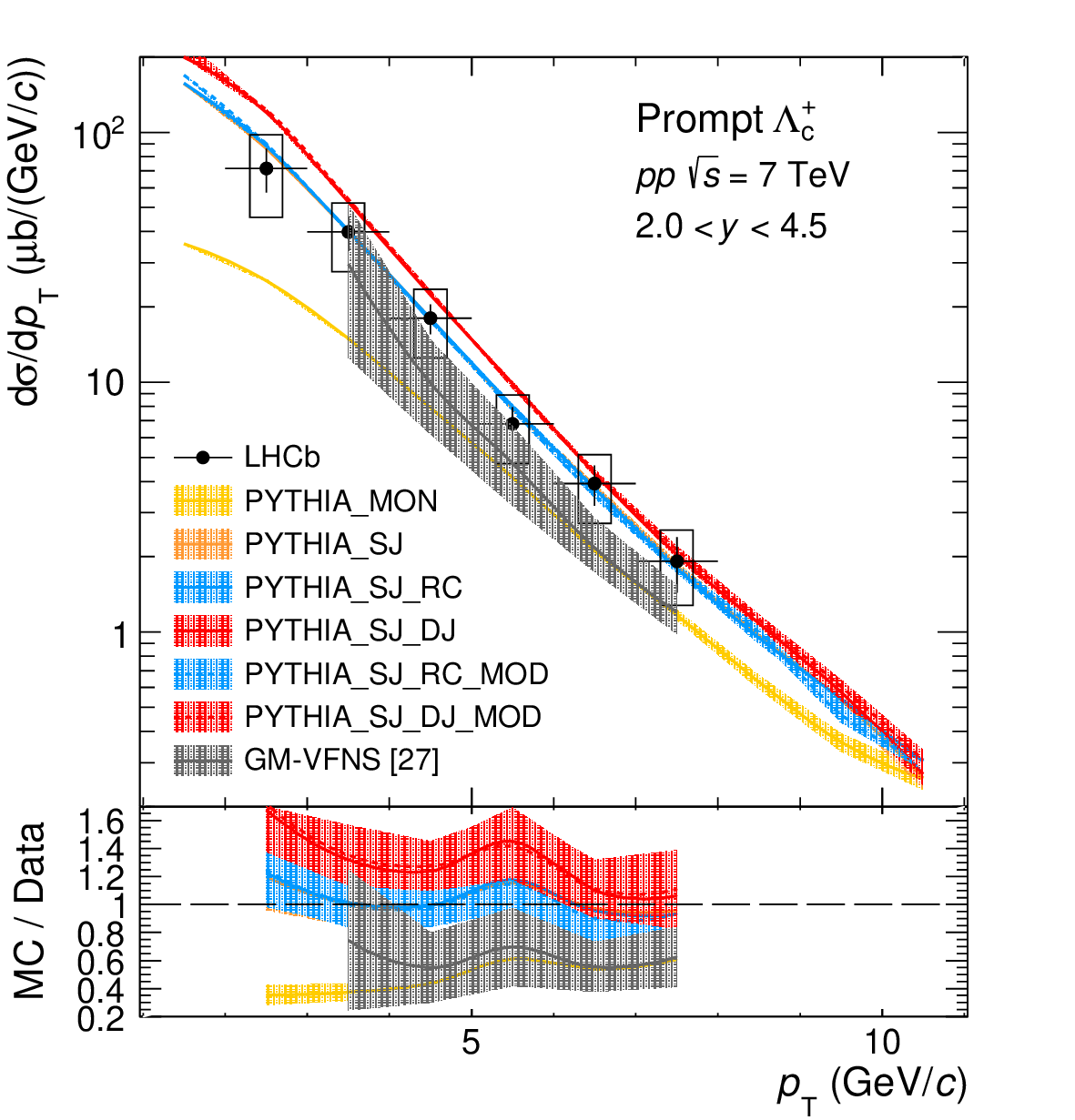}
    \label{fig:lc_pt_lhcb}
    
  }
  \subfigure[]{
    \includegraphics[width=0.45\linewidth]{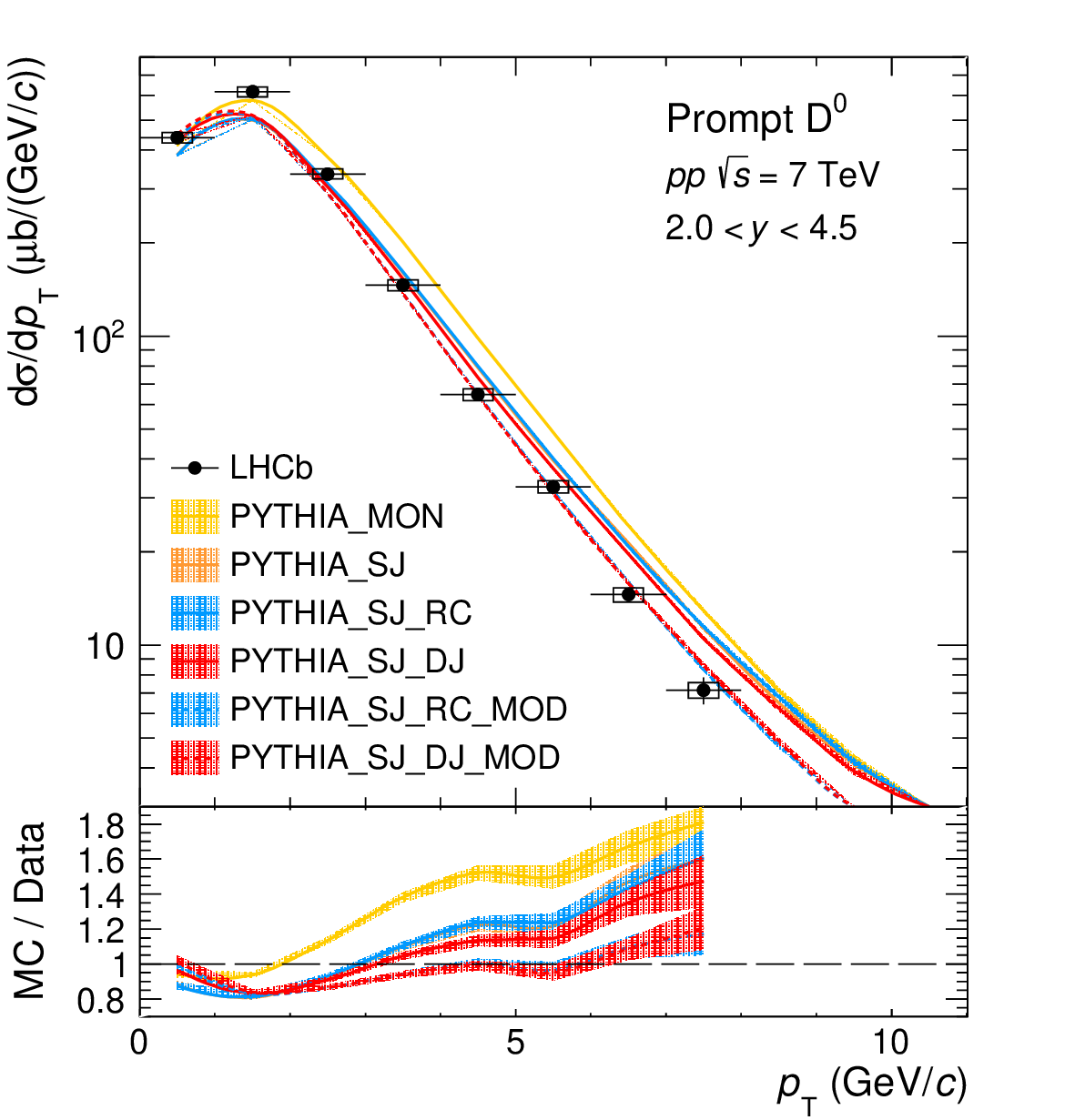}
    \label{fig:d0_pt_lhcb}
  } 
  \hfill
  \caption{$p_\mathrm{T}$-dependent differential production cross-section of $\Lambda_c^+$ [(a), (c)] and $D^0$ [(b), (d)] measured by ALICE \cite{2018, PhysRevC.94.054908} and LHCb \cite{20131} experiments in p+p collisions at $\sqrt{s}=$ 7 TeV respectively, compared with various model predictions. The error bars associated with the data points represent the statistical uncertainty, whereas the square boxes around the data points represent the systematic uncertainty. The shaded region around the solid and dashed lines represents the uncertainty associated with the models. The lower panel of each plot illustrates the ratio between the model and the experimental data points.}
  \label{fig:c_diffprod}
\end{figure*}

The Fig. \ref{fig:c_diffprod} shows the ALICE and LHCb measurements on the $p_\mathrm{T}$-dependent differential production cross-section of prompt  $\Lambda_c^+$ and $D^0$, compared with the predictions from the considered tunes of the {\large P}YTHIA generated datasets of the present investigation. In these figures, the black markers represent the experimental data points of ALICE and LHCb, with the corresponding error bars indicating statistical uncertainties. The boxes surrounding these markers represent systematic uncertainties associated with the measurements. The lower panels of each plot show the ratios between the various model-generated predictions and the corresponding experimental data points. From Fig.\ref{fig:lc_pt_alice} - \ref{fig:d0_pt_lhcb}, it is evident that the default {\large P}YTHIA Monash tune ({\large P}YTHIA\_MON) greatly underestimates the production of $\Lambda_c^+$ at both ALICE and LHCb acceptances. Conversely, it overestimates the production of $D^0$, particularly at high $p_\mathrm{T}$, for both these experiments. 
Within the acceptance of ALICE, though {\large P}YTHIA\_SJ and {\large P}YTHIA\_SJ\_RC, with the default color reconnection mode-1, could result in a better estimation of $\Lambda_c^+$ and $D^0$ production than the estimated values of the earlier models, including {\large P}YTHIA\_MON, there is a noticeable difference in the predictions of the model yield of $\Lambda_c^+$ in particular, implying that the string junction alone with the default color reconnection mode-1 is not enough to describe the ALICE results of $\Lambda_c^+$ production. 
However, within the acceptance of LHCb, the same two datasets {\large P}YTHIA\_SJ and {\large P}YTHIA\_SJ\_RC show a better agreement with the experimental yield of $\Lambda_c^+$ and $D^0$ (Fig.\ref{fig:lc_pt_lhcb} and \ref{fig:d0_pt_lhcb}). Within ALICE and LHCb acceptances, the datasets {\large P}YTHIA\_SJ and {\large P}YTHIA\_SJ\_RC almost coincide with each other, indicating that 'liberated' string junctions from beam remnants do not play any significant role in the production of $\Lambda_c^+$ and $D^0$.

Furthermore, it has been observed that both models overestimate the $p_\mathrm{T}$-differential production cross-section for the $D^0$ meson compared to the experimental measurements from ALICE and LHCb. Since mesons are primarily produced through string fragmentation, as described by the Lund String model \cite{10.21468/SciPostPhysCodeb.8, Andersson1983-kb, SJOSTRAND1984469}, this discrepancy is believed to have arisen from the parametrization of the Lund symmetric fragmentation function \cite{Skands2014-rk}, which results in increased yields at high $p_\mathrm{T}$. The expression of this symmetric fragmentation function is given as follows \cite{10.21468/SciPostPhysCodeb.8},
\begin{equation} \label{eq:Lund_sff}
    f(z) \propto \frac{(1-z)^a}{z} \exp{\left(-\frac{bm_T^2}{z} \right)}
\end{equation}
where $z$ is the fraction of longitudinal momentum that a newly formed hadron takes from the remaining string system during a single string fragmentation, and $m_T$ denotes the transverse mass. \\
The parameters $a$ and $b$ are known as Lund parameters, which characterize the shape of the fragmentation function and must be constrained by a fit to experimental data \cite{Skands2014-rk}. By adjusting these parameters independently, one can alter both the average hardness and the width of the fragmentation spectra, thus impacting the resultant $p_\mathrm{T}$ spectra. In {\large P}YTHIA-8.314, the default values for these parameters are set to \emph{a = 0.68} and \emph{b = 0.98}, which are based on the fits to $e^+e^-$ data \cite{Skands2014-rk}. However, the expression in Eq.\ref{eq:Lund_sff} is only valid for string ends attached to light-flavored quarks (antiquarks), which can be treated as essentially massless \cite{10.21468/SciPostPhysCodeb.8}. In cases where one of the string ends is attached to a heavier quark, such as a charm or bottom quark, the Bowler's modification \cite{Bowler1981-sj} incorporated in the Artru-Mennesier model \cite{ARTRU197493} is generally preferred over the Lund symmetric fragmentation function. This modified symmetric fragmentation function is given as,
\begin{equation}
    f(z) \propto \frac{(1-z)^a}{z^{1+r_Q bm_Q^2}} \exp{\left(-\frac{bm_T^2}{z} \right)}
\end{equation}
where, $m_Q$ represents the mass of the heavy quark, while $r_Q$ is a multiplicative factor, typically approximated to unity but can be constrained by a fit to experimental data.

\begin{figure*}
  \subfigure[]{
    \includegraphics[width=0.45\linewidth]{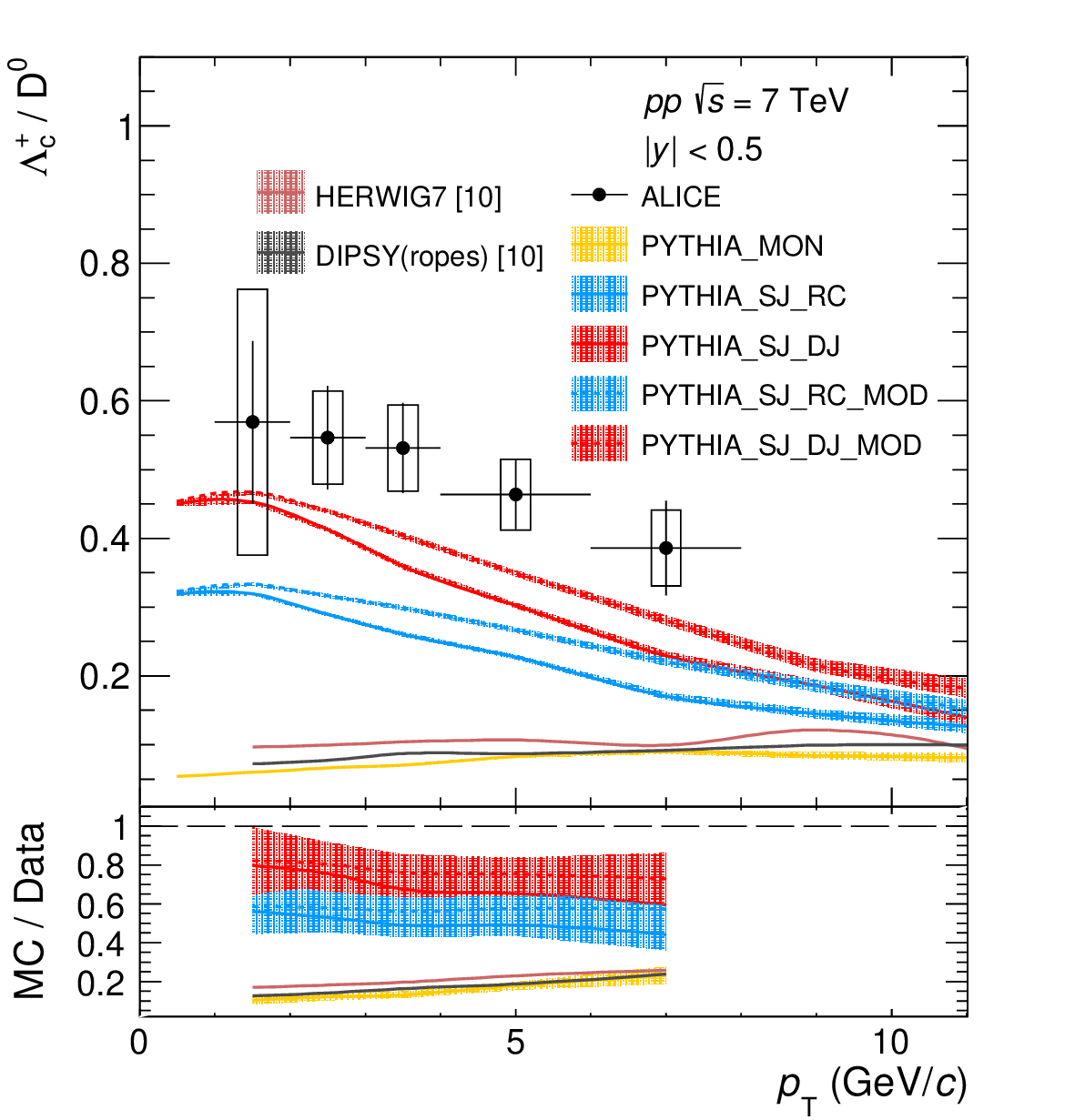}
    \label{fig:lcd0ratio_pt_alice}
    
  }
  \subfigure[]{
    \includegraphics[width=0.45\linewidth]{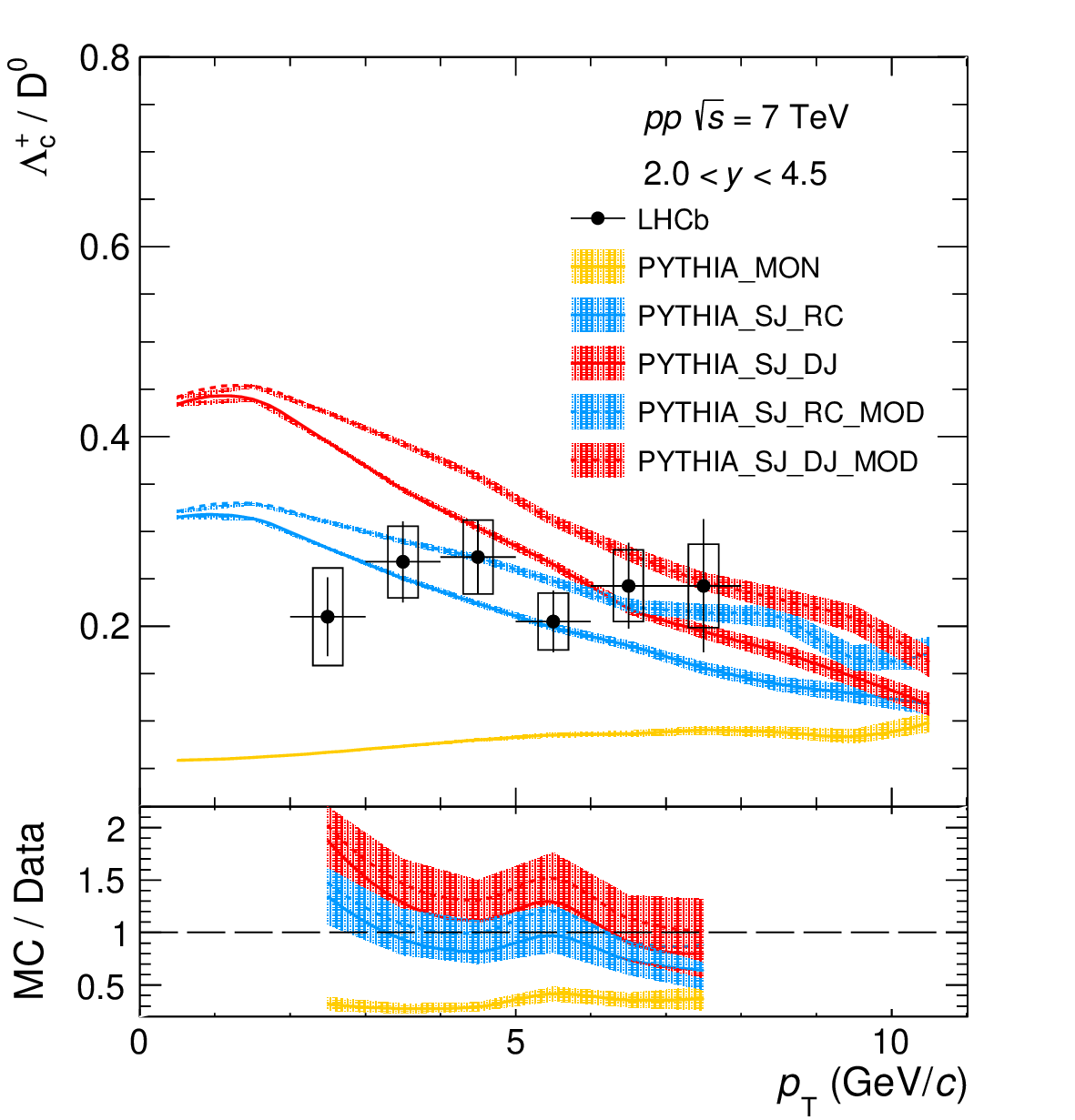}
    \label{fig:lcd0ratio_pt_lhcb}
  } 
  \hfill
  \caption{Comparisons of $p_\mathrm{T}$-dependent $\Lambda_c^+/D^0$ ratio measured by (a) ALICE\cite{2018} and (b) LHCb\cite{2018, 20131} in p+p collisions at $\sqrt{s}=$ 7 TeV with different model predictions. The error bars associated with the data-points represent the statistical uncertainty, whereas the square boxes around the data points represent the systematic uncertainty.}
  \label{fig:c_ratio}
\end{figure*}

Altmann et al., in their recent publication \cite{Altmann2024-wm}, reported their findings on the estimation of the $\Lambda_c^+/D^0$ ratio using MC data for p+p collisions at $\sqrt{s}=$ 13 TeV. Their study demonstrated that the string junction model in the QCDCR scheme (color Reconnection mode - 1) of {\large P}YTHIA, exhibited an improved agreement with the experimental measurement from ALICE \cite{PhysRevLett.128.012001} on the $\Lambda_c^+/D^0$ ratio. Notably, the authors used non-default values for several tuning parameters in their generated set of data, including Lund parameters and Bowler's multiplicative factor set to $a=0.36$, $b=0.56$, and $r_c=1.5$. Additionally, they disabled in their generated data set, a flag \verb|ColourReconnection:AllowDoubleJunRem|, which is associated with the QCD-based color reconnection scheme. The complete set of tuning parameters used in their investigation is detailed in the Appendix section of the ref. \cite{Altmann2024-wm}.
Since the work of Altmann et al. on the $\Lambda_c^+/D^0$ ratio was limited to the ALICE results on p+p collisions at 13 TeV, we generated a new {\large P}YTHIA dataset for p+p collisions at $\sqrt{s} = 7$ TeV, incorporating the parameters specified in the ref. \cite{Altmann2024-wm}. The objective of this exercise is to investigate the extent of agreement between the results of experimental measurements of ALICE and LHCb on the production cross section of $\Lambda_c^+$ and $D^0$ for p+p collisions at $\sqrt{s}=$ 7 TeV with the above-mentioned non-default implementations of the QCDCR model. This newly generated dataset is labeled as {\large P}YTHIA\_SJ\_DJ. \\
It can be easily seen from the same Fig.\ref{fig:lc_pt_alice} that the plot with this newly generated dataset ({\large P}YTHIA\_SJ\_DJ) is in much better agreement with the ALICE results on $\Lambda_c^+$ production, surpassing previous model comparisons. Moreover, it also reduces the previously observed discrepancies in the $D^0$ production compared to the ALICE and LHCb measurements. This improved agreement with the $D^0$ production, compared to the default QCDCR model, is correlated with the adjusted parameters associated with the symmetric fragmentation function implemented in this dataset. The better agreement is similarly reflected in Fig. \ref{fig:lcd0ratio_pt_alice}, where the result of the generated dataset (PYTHIA\_SJ\_DJ) shows a better agreement with the ALICE experimental results on the $\Lambda_c^+/D^0$ ratio measurement. The observed enhancement in the production of the $\Lambda_c^+$ baryon, and thus the $\Lambda_c^+/D^0$ ratio, may be associated with the deactivation of the flag \verb|ColourReconnection:AllowDoubleJunRem|. This flag, integral to the QCDCR scheme in {\large P}YTHIA, governs whether a directly connected junction-antijunction pair can be reconfigured into two distinct string configurations. The final outcome is based on comparing the $\lambda$-measure (or the string length) \cite{Christiansen2015-dt} of the junction antijunction pair with that of the two possible string configurations. When the flag is disabled, the junction system is preserved, and a $q-\bar{q}$ pair may be inserted between the junction-antijunction pair, with energy and momentum redistributed from the adjacent legs of the junction system \cite{10.21468/SciPostPhysCodeb.8, Christiansen2015-dt}. This mechanism effectively increases the likelihood of baryon and antibaryon production, such as $\Lambda_c^+$ and its antiparticle, during the hadronization process. However, despite these modifications, the production of $\Lambda_c^+$ and thus, the $\Lambda_c^+/D^0$ ratio remains still underestimated relative to ALICE measurements at $\sqrt{s} = $ 7 TeV. On the other hand, the increased production of $\Lambda_c^+$ baryons adversely affects the model's agreement with the LHCb measurements at $\sqrt{s}=$ 7 TeV, as demonstrated in Fig. \ref{fig:lc_pt_lhcb}. As a result, the model tends to significantly overestimate the $\Lambda_c^+/D^0$ ratio reported by LHCb (Fig.\ref{fig:lcd0ratio_pt_lhcb}), although achieving improved agreement with the LHCb measured $D^0$ production.

Thus, while the tuning parameters associated with the symmetric fragmentation function within {\large P}YTHIA\_SJ\_DJ (for $\sqrt{s} = $ 7 TeV) have reduced the discrepancies previously observed with our earlier {\large P}YTHIA\_SJ\_RC model estimation of $D^0$ production for both ALICE and LHCb measurements, some disagreement still persists. To address these, we have embarked on further refinement of these tuning parameters, with the aim of improving the agreement between the experimental and MC estimated $D^0$ meson production for both ALICE and LHCb. For this tuning procedure, we began with the default implementations of the QCDCR model of {\large P}YTHIA, iteratively adjusting the parameters related to the symmetric fragmentation function using the hit and trial method, until we reached a minimum value of $\chi^2/\text{NDF}$ between the prediction of the model and the experimental data, and is listed in Table\ref{table:tuning_chi2ndf}. The values for the Lund parameters ($a$ and $b$), alongside Bowler's multiplicative factor for charm quark ($r_c$), for which we settle in the present investigation are, \emph{a = 1.1}, \emph{b = 0.6 }, and \emph{$r_c =$ 1.5}. 

\begin{table}
\caption{\label{table:tuning_chi2ndf}Minimum values of $\chi^2/\text{NDF}$ as obtained from various model comparisons with the experimental $p_\mathrm{T}$-differential production cross-section of $D^0$ mesons}
\begin{ruledtabular}
\begin{tabular}{l c c} 
Datasets& $D^0$ ALICE & $D^0$ LHCb \\
\hline
{\large P}YTHIA\_MON& 9.67714& 112.598\\
\hline
{\large P}YTHIA\_SJ\_DJ&  5.4171& 38.8258\\
\hline
{\large P}YTHIA\_SJ\_RC& 8.68618& 76.3519\\
\hline
{\large P}YTHIA\_SJ\_RC with& &\\
$a = 1.1$, $b = 0.6$, $r_C = 1.5$& 2.31628 & 13.1008
\end{tabular}
\end{ruledtabular}
\end{table}

The rationale for these specific tunings is to reduce the high $p_\mathrm{T}$ yield of the $D^0$ meson, and thus, minimize the discrepancies in comparison to experimental observations. However, these tunings also exert an influence on the high $p_\mathrm{T}$ yield of $\Lambda_c^+$ baryons, as they affect the momentum fractions acquired by all hadrons formed through the string fragmentation process \cite{Skands2014-rk}.
Guided by the fact that in PYTHIA, an enhanced diquark relative to quark production enhances the baryon relative to meson production \cite{Andersson1982-aq}, to compensate for underproduction of $\Lambda_c^+$ baryons at high $p_\mathrm{T}$, the diquark production relative to quark production \cite{Skands2014-rk} in the string fragmentation process was incrementally increased from its default value of 0.081 to 0.12. In the {\large P}YTHIA event generator, these adjustments are configured as follows,\\
\verb| StringZ:aLund=1.1|\\
\verb| StringZ:bLund=0.6|\\
\verb| StringZ:rFactC=1.5|\\
\verb| StringFlav:probQQtoQ=0.12|

\begin{figure}
  \includegraphics[width=\linewidth]{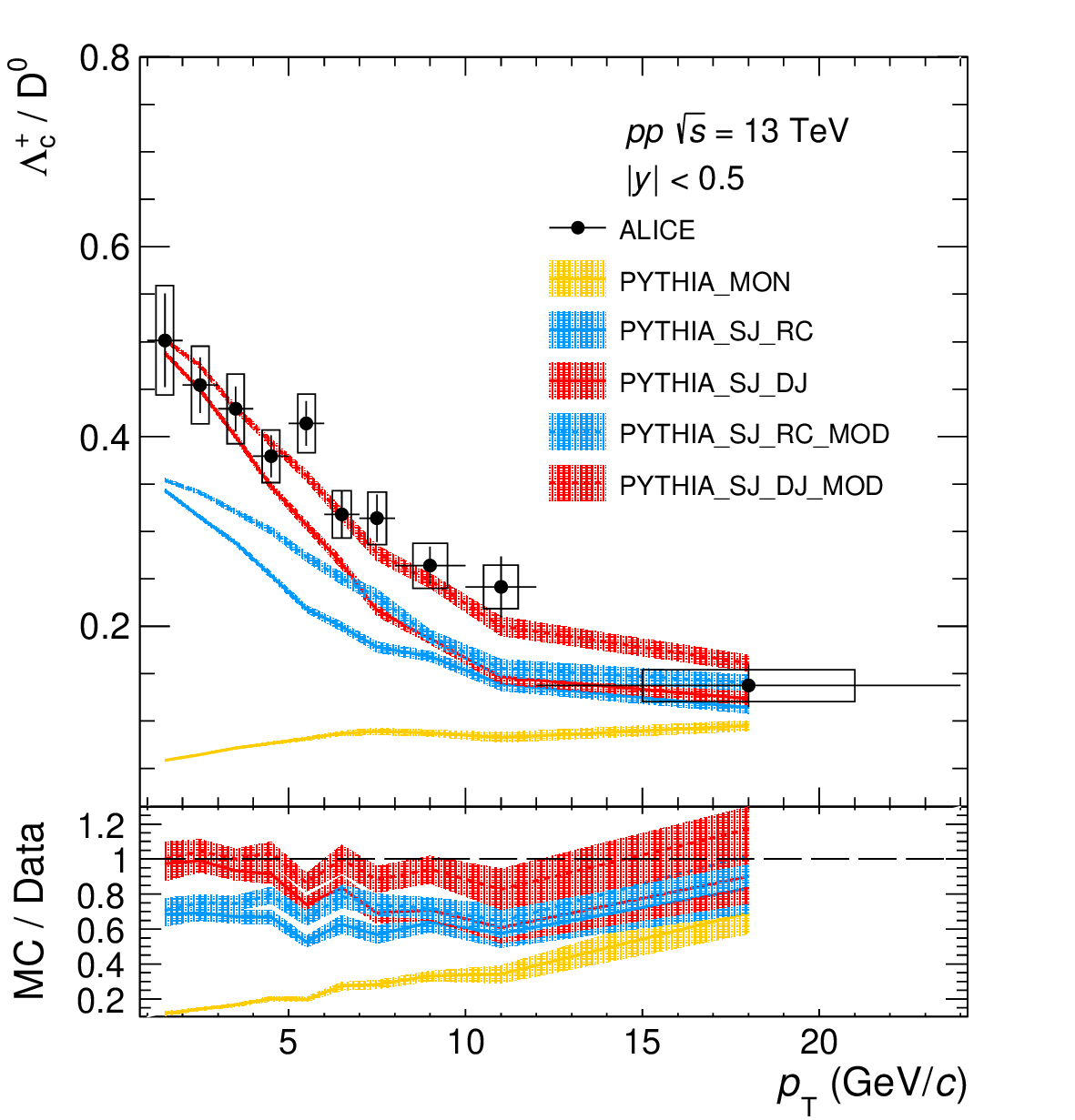}
  \caption{Comparison of  $p_\mathrm{T}$-dependent $\Lambda_c^+/D^0$ ratio measured by ALICE \cite{PhysRevLett.128.012001} in p+p collisions at $\sqrt{s}=$ 13 TeV with different model predictions. The error bars associated with the data-points represent the statistical uncertainty, whereas the square boxes around the data points represent the systematic uncertainty.}
  \label{fig:lcd0ratio_pt_alice_13tev}
\end{figure} 
Subsequently, the {\large P}YTHIA\_SJ\_RC and the {\large P}YTHIA\_SJ\_DJ datasets have been regenerated, incorporating these updated parameters for both $\sqrt{s} = $ 7 and 13 TeV. For ease of reference, these two newly generated datasets will be labeled as {\large P}YTHIA\_SJ\_RC\_MOD and {\large P}YTHIA\_SJ\_DJ\_MOD. It is important to note that we have utilized the {\large P}YTHIA\_SJ\_RC instead of the {\large P}YTHIA\_SJ, considering the demonstrated efficacy of the former as elaborated in ref.\cite{PhysRevC.111.014905}.\\
The comparisons of the outcomes from these newly generated datasets ({\large P}YTHIA\_SJ\_RC\_MOD and {\large P}YTHIA\_SJ\_DJ\_MOD) with the experimental measurements are shown in the Figures \ref{fig:c_diffprod} and \ref{fig:c_ratio}. The plots in Fig. \ref{fig:c_diffprod} show that with these modifications, both {\large P}YTHIA\_SJ\_RC\_MOD and {\large P}YTHIA\_SJ\_DJ\_MOD exhibit the most substantial agreement with the experimentally measured $p_\mathrm{T}$-dependent $D^0$ production cross-section, as reported by ALICE and LHCb. Additionally, these modifications lead to a marked enhancement in the production of $\Lambda_c^+$ baryons in {\large P}YTHIA\_SJ\_DJ\_MOD, thereby increasing the $\Lambda_c^+/D^0$ ratio, which aligns more closely with the measurements from ALICE for both $\sqrt{s}=$ 7 (Fig.\ref{fig:lcd0ratio_pt_alice}) and 13 TeV results (Fig.\ref{fig:lcd0ratio_pt_alice_13tev}). This enhancement, nonetheless, results in an overestimation when compared to LHCb measurements (Fig.\ref{fig:lcd0ratio_pt_lhcb}). However, a certain degree of underestimation still persists in the alignment of the generated data-points with the $\sqrt{s}=$ 7 TeV ALICE experimental data-points. 

 In contrast, {\large P}YTHIA\_SJ\_RC\_MOD demonstrates a commendable capability to accurately reproduce the LHCb measurements on the $\Lambda_c^+/D^0$ ratio (Fig.\ref{fig:lcd0ratio_pt_lhcb}). Within the ALICE detector acceptance, this model-generated dataset shows a marginal improvement in the agreement with the ALICE measurements compared to the default string junction model implementation (Fig.\ref{fig:lcd0ratio_pt_alice} and \ref{fig:lcd0ratio_pt_alice_13tev}).

In conclusion, the comparison of the $p_\mathrm{T}$-differential production cross-sections of $\Lambda_c^+$ and $D^0$, along with their respective ratio, as measured by both ALICE and LHCb experiments with our MC results, indicates that the {\large P}YTHIA\_SJ\_DJ\_MOD generated datasets of this investigation gives the best agreement with the experimental results of ALICE on $p_\mathrm{T}$-differential production cross-section of $\Lambda_c^+$, $D^0$ as well as $p_\mathrm{T}$-dependent $\Lambda_c^+/D^0$ ratio among all the studied models. However, the experimental results of LHCb on all the above-mentioned observables are found to be best represented by our generated dataset of {\large P}YTHIA\_SJ\_RC\_MOD model.

\subsection{\label{subsec:results_bottom}Comparison of \textbf{$\Lambda_b^0$} and \textbf{$\bar{B}^0(B^0)$} experimental results with model estimations}
\begin{figure*}
  \subfigure[]{
    \includegraphics[width=0.45\linewidth]{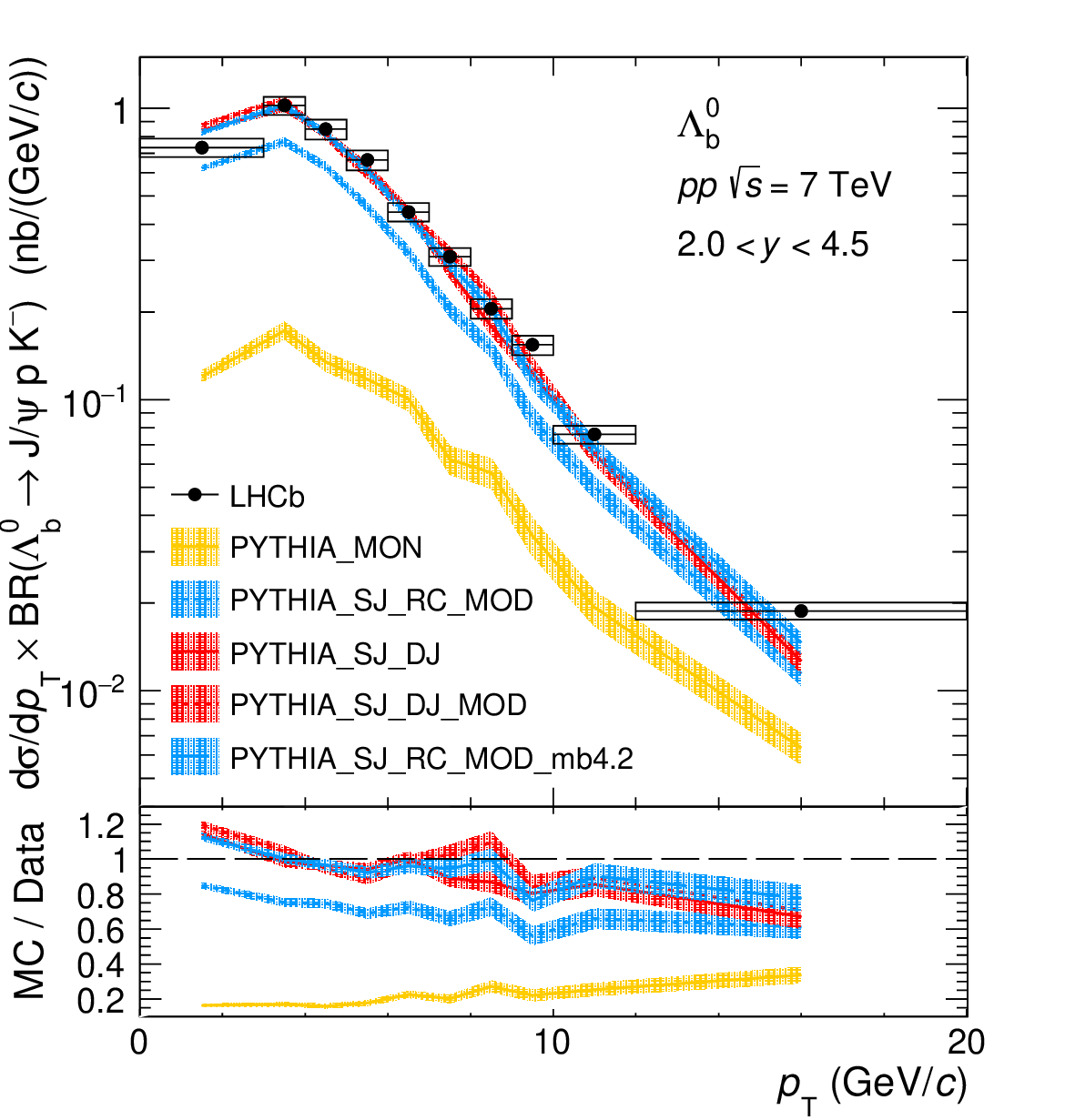}
    \label{fig:l0b_pt_lhcb}
    
  }
  \subfigure[]{
    \includegraphics[width=0.45\linewidth]{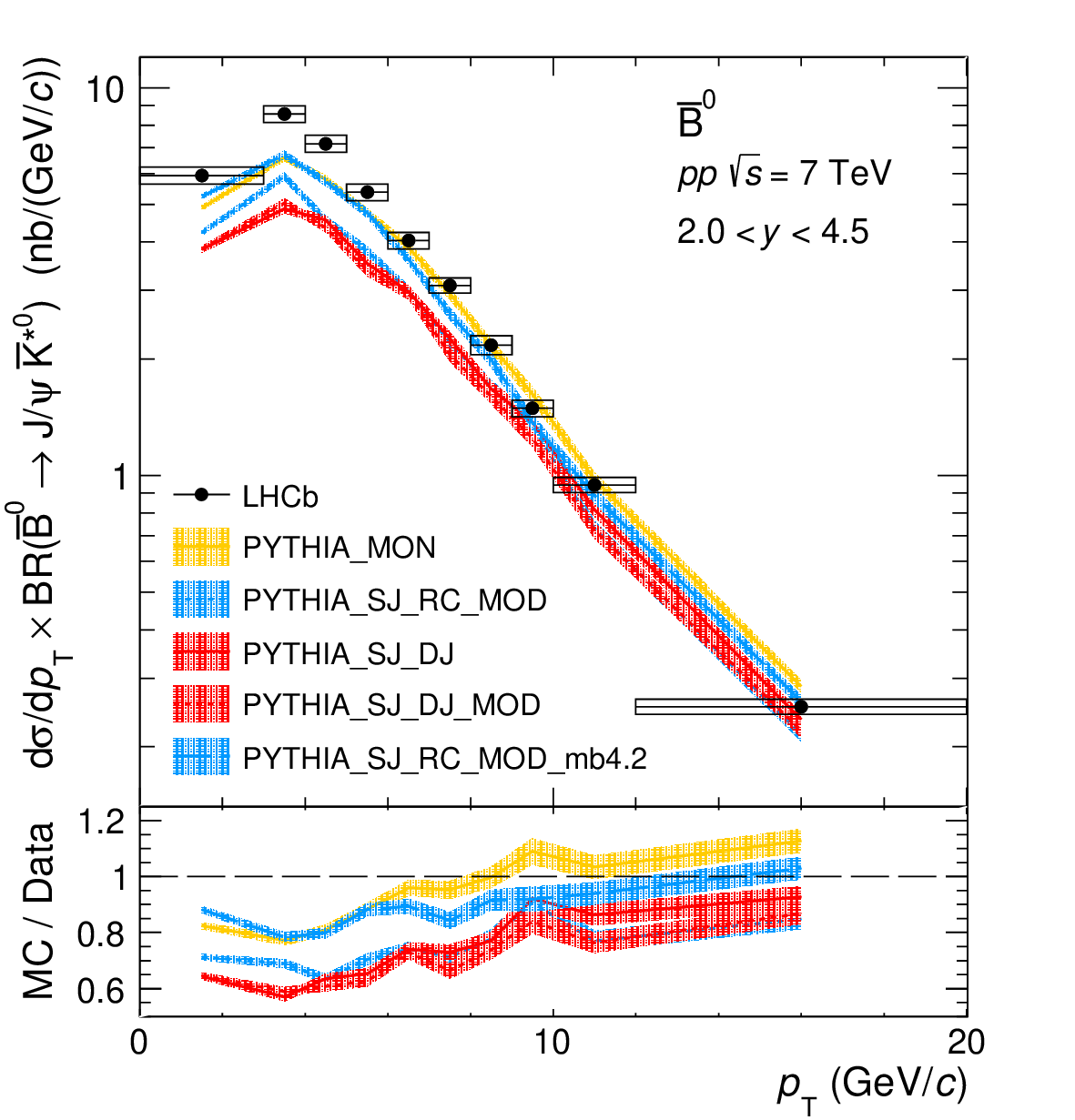}
    \label{fig:ab0_pt_lhcb}
  } 
  \hfill
  \caption{A comparison of model predictions with the experimental $p_\mathrm{T}$-dependent differential production cross-section times branching ratio of (a) $\Lambda_b^0$ and (b) $\bar{B}^0$ measured by LHCb \cite{Aaij_2016} in p+p collisions at $\sqrt{s} = $ 7 TeV. The error bars associated with the data-points represent the statistical uncertainty, whereas the square boxes around the data points represent the systematic uncertainty.}
  \label{fig:b_diffprod}
\end{figure*} 

\begin{figure*}
  \subfigure[]{
    \includegraphics[width=0.45\linewidth]{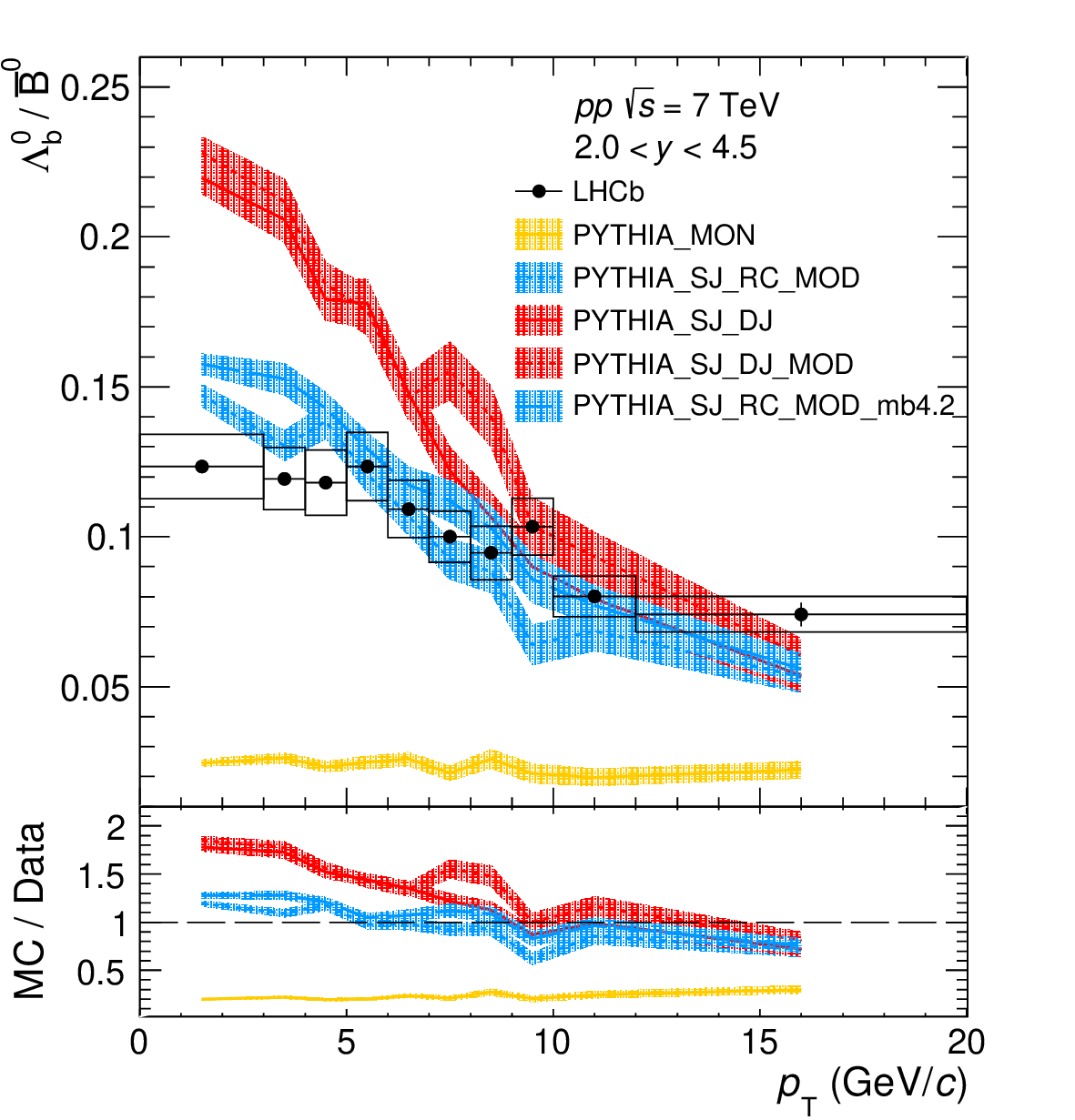}
    \label{fig:l0bab0ratio_pt_lhcb}
    
  }
  \subfigure[]{
    \includegraphics[width=0.45\linewidth]{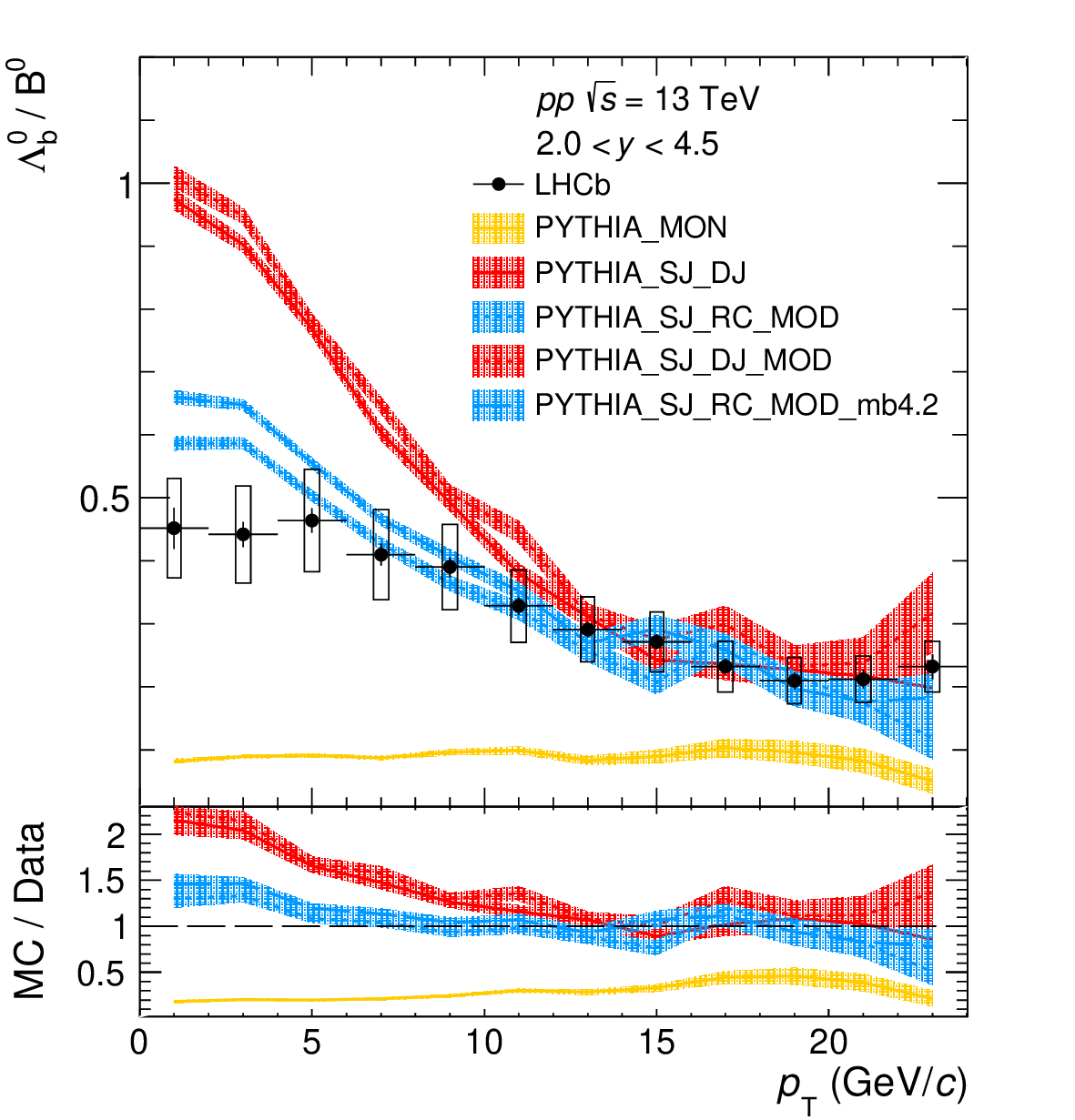}
    \label{fig:L0b_B0_ratio _13TeV}
  } 
  \hfill
  \caption{A comparison of model predictions with experimental $p_\mathrm{T}$-dependent (a) $\Lambda_b^0/\bar{B}^0$ and (b) $\Lambda_b^0/B^0$ ratio measured by LHCb \cite{Aaij_2016, PhysRevLett.132.081901} in p+p collisions at $\sqrt{s}=$ 7 TeV and $\sqrt{s}=$ 13 TeV. The error bars associated with the data-points represent the statistical uncertainty, whereas the square boxes around the data points represent the systematic uncertainty.}
  \label{fig:b_ratio}
\end{figure*} 
Altmann et al. \cite{Altmann2024-wm}, using the same model (here labeled as {\large P}YTHIA\_SJ\_DJ) for p+p collisions at $\sqrt{s}=13$ TeV, which they used to describe the ALICE result on the $\Lambda_c^+/D^0$ ratio at the same energy, also estimated $\Lambda_b^0/B^0$ ratio and compared their result with the available LHCb experimental result \cite{PhysRevLett.132.081901}. Their findings showed that {\large P}YTHIA\_SJ\_DJ model generated data significantly overestimated the LHCb results. Moreover, there are reports from the LHCb experiment on the $p_\mathrm{T}$-differential production cross-section times the branching ratio of $\Lambda_b^0$, $\bar{B}^0$ and their ratio, for p+p collisions at $\sqrt{s}=7$ TeV \cite{Aaij_2016}. However, no attempt has been made so far to make a comparison of the above-mentioned LHCb experimental findings with the model prediction. In this section, we make an attempt to discuss these experimental results with {\large P}YTHIA\_SJ\_RC\_MOD generated data. 
It is important to note that charge conjugation was assumed for the LHCb result on the $\Lambda_b^0/B^0$ ratio at $\sqrt{s}=$ 13 TeV \cite{PhysRevLett.132.081901}, suggesting that our investigation into the production of $\bar{B}^0$ instead of $B^0$ should remain unaffected.
Our {\large P}YTHIA\_SJ\_RC\_MOD  model predicted results, along with the other model predictions on estimation of the $p_\mathrm{T}$-differential production cross-section times the branching ratio of $\Lambda_b^0$ and $\bar{B}^0$ for p+p collisions at $\sqrt{s}=$ 7 TeV, as well as the LHCb measured results, are shown in Fig.\ref{fig:b_diffprod}.\\
The plots presented in Fig.\ref{fig:b_diffprod} clearly show that, within LHCb acceptance, the {\large P}YTHIA\_MON model is capable of describing the production of $\bar{B}^0$ mesons; however, it significantly underestimates the production of $\Lambda_b^0$ baryons. On the other hand, {\large P}YTHIA\_SJ\_DJ, along with its modified dataset {\large P}YTHIA\_SJ\_DJ\_MOD, which incorporates the adjustments on the parametrization of the symmetric fragmentation function, though demonstrates a more favorable estimation for $\Lambda_b^0$ production (Fig.\ref{fig:l0b_pt_lhcb}), concurrently underestimates the $\bar{B}^0$ production (Fig.\ref{fig:ab0_pt_lhcb}). This observed discrepancy may be attributed to the deactivation of the flag \verb|ColourReconnection:allowDoubleJunRem| in the {\large P}YTHIA\_SJ\_DJ (as well as {\large P}YTHIA\_SJ\_DJ\_MOD). Such deactivation leads to an increase in the production of $\Lambda_b^0$  baryons than the production of $\bar{B}^0$ mesons. The reduced yield of  $\bar{B}^0$ mesons results in a considerable overestimation of the $\Lambda_b^0/\bar{B}^0$ ratio at $\sqrt{s}=$ 7 TeV (Fig.\ref{fig:l0bab0ratio_pt_lhcb}). A similar conclusion can be drawn regarding the discrepancy observed by Altmann et al. \cite{Altmann2024-wm} on $\Lambda_b^0/B^0$ ratio at $\sqrt{s}=$ 13 TeV (Fig.\ref{fig:L0b_B0_ratio _13TeV}). On the other hand, {\large P}YTHIA\_SJ\_RC\_MOD underestimates the production of both $\Lambda_b^0$ and $\bar{B}^0$ by a comparable margin, suggesting that the string junction model in the default CR mode-1 may not adequately account for the production of bottom quarks. This inadequacy is supported by the results shown in Fig. \ref{fig:l0b_pt_lhcb} and \ref{fig:ab0_pt_lhcb}.

Given that the mass of the bottom quark (as well as charm quark) is significantly greater than the $\Lambda_{QCD}$ scale, they cannot be produced through non-perturbative hadronisation processes, such as via string fragmentation \cite{Altmann2025-mj, Altmann2024-wm, Norrbin2000-ij}. Instead, these quarks must be produced before the hadronisation during the initial hard scatterings through leading-order perturbative processes, such as $q\bar{q} \longrightarrow b\bar{b}$, $gg\longrightarrow b\bar{b}$, as well as through parton showers involving gluon splitting \cite{Norrbin2000-ij,Bierlich2024-kq}. The mass of the bottom quark is a critical parameter in the perturbative description, influencing both the matrix element calculations and the available phase space \cite{Norrbin2000-ij}. Therefore, the chosen value for the bottom quark mass exhibits considerable sensitivity concerning the production cross-section of bottom quarks. In {\large P}YTHIA version 8.314, the default mass of the bottom quark is set as $m_b = 4.8 \, GeV/c^2$, which is based on a standardized mass for charm quark ($m_c = 1.5 \, GeV/c^2$) \cite{NORRBIN1998407} and the conventional mass formula referenced in \cite{PhysRevD.12.147, Norrbin2000-ij}.

In pursuit of increasing the production cross-section of bottom quarks, the mass was reduced to a value of $m_b = 4.2 \, GeV/c^2$ (via \verb|5:m0=4.2|), as proposed in ref.\cite{Bierlich2024-kq}. This configuration lowers the threshold required for the creation of bottom quark and antiquark pairs during both initial hard scatterings and parton showers \cite{Norrbin2000-ij}, thereby facilitating an increased production cross-section for bottom hadrons. Additionally, similar to the modifications implemented for charm quarks in sub-section (\ref{subsec:results_charm}), the Bowler's multiplicative factor for bottom quark ($r_b$) is raised from its default of 0.855 to 0.96 again by the hit and trial method. 
With these modified values of $m_b$ and $r_b$, keeping all other parameters the same, the dataset {\large P}YTHIA\_SJ\_RC\_MOD is regenerated, and this newly generated dataset is named for convenience as {\large P}YTHIA\_SJ\_RC\_MOD\_mb4.2. The tuning parameters associated with this dataset are summarized in Table-\ref{table:tuningpar}. As evident from the Fig.\ref{fig:b_diffprod}, this dataset provides the most favorable agreement with the $p_\mathrm{T}$-dependent differential production cross-section of $\Lambda_b^0$ and $\bar{B}^0$, indicating that the default bottom quark mass in {\large P}YTHIA may be high enough to accurately reflect the substantial production of bottom quarks as observed experimentally.
\begin{table*}[!htbp]
\centering
\caption{\label{table:tuningpar}Full list of parameters of the {\large P}YTHIA-8.314 generated datasets}
\resizebox{\textwidth}{!}{%
\begin{ruledtabular}
\begin{tabular}{l|ccccccc}
Parameters &
\rotatebox{70}{PYTHIA\_MON (default)} &
\rotatebox{70}{PYTHIA\_SJ} &
\rotatebox{70}{PYTHIA\_SJ\_RC} &
\rotatebox{70}{PYTHIA\_SJ\_RC\_MOD} &
\rotatebox{70}{PYTHIA\_SJ\_RC\_MOD\_mb4.2} &
\rotatebox{70}{PYTHIA\_SJ\_DJ} &
\rotatebox{70}{PYTHIA\_SJ\_DJ\_MOD} \\
\hline
ColourReconnection:mode              & 0     & 1   & 1   & 1   & 1   & 1   & 1 \\
ColourReconnection:allowDoubleJunRem & on    & on  & on  & on  & on  & off & off \\
BeamRemnants:remnantMode             & 0     & 1   & 1   & 1   & 1   & 1   & 1 \\
BeamRemnants:beamJunction            & --    & off & on  & on  & on  & off & off \\
StringZ:aLund                        & 0.68  & 0.68& 0.68& 1.1 & 1.1 & 0.36& 1.1 \\
StringZ:bLund                        & 0.98  & 0.98& 0.98& 0.6 & 0.6 & 0.56& 0.6 \\
StringZ:rFactC                       & 1.32  & 1.32& 1.32& 1.5 & 1.5 & 1.5 & 1.5 \\
StringZ:rFactB                       & 0.855 &0.855&0.855&0.855&0.96&0.855&0.855 \\
StringFlav:mesonCvector              & 0.88  & 0.88& 0.88& 0.88& 0.88& 1.5 & 1.5 \\
StringFlav:probStoUD                 & 0.217 &0.217&0.217&0.217&0.217&0.20&0.20 \\
StringFlav:probQQtoQ                 & 0.081 &0.081&0.081&0.12 &0.12 &0.078&0.12 \\
MultiPartonInteractions:pT0Ref       & 2.28  & 2.28& 2.28& 2.28& 2.28& 2.25& 2.25 \\
5:m0                                 & 4.8   & 4.8 & 4.8 & 4.8 & 4.2 & 4.8 & 4.8 \\
\end{tabular}
\end{ruledtabular}
}
\end{table*}
From Fig. \ref{fig:l0bab0ratio_pt_lhcb}, it is interesting to see that despite {\large P}YTHIA\_SJ\_RC\_MOD underestimates the production of both $\Lambda_b^0$ and $\bar{B}^0$, it still exhibits reasonably good agreement with the experimentally measured $\Lambda_b^0 / \bar{B}^0$ ratio. However, the dataset {\large P}YTHIA\_SJ\_RC\_MOD\_mb4.2 not only gives good agreement with the experimentally measured values of $p_\mathrm{T}$-differential production cross-section of $\Lambda_b^0$, $\bar{B}^0$ but also $\Lambda_b^0/\bar{B}^0$ ratio at $\sqrt{s} =$ 7 (Fig.\ref{fig:l0bab0ratio_pt_lhcb}), and $\Lambda_b^0/B^0$ at $\sqrt{s} =$ 13 TeV (Fig.\ref{fig:L0b_B0_ratio _13TeV}), indicating that {\large P}YTHIA\_SJ\_RC\_MOD\_mb4.2 remains the most successful model out of the all the considered models of the present investigation in describing the considered heavy baryon-to-heavy meson ratios at LHCb acceptance.

\subsection{\label{subsec:results_robustness_check}Robustness check of {\large P}YTHIA\_SJ\_RC\_MOD\_mb4.2 model}

\begin{figure}
  \includegraphics[width=\linewidth]{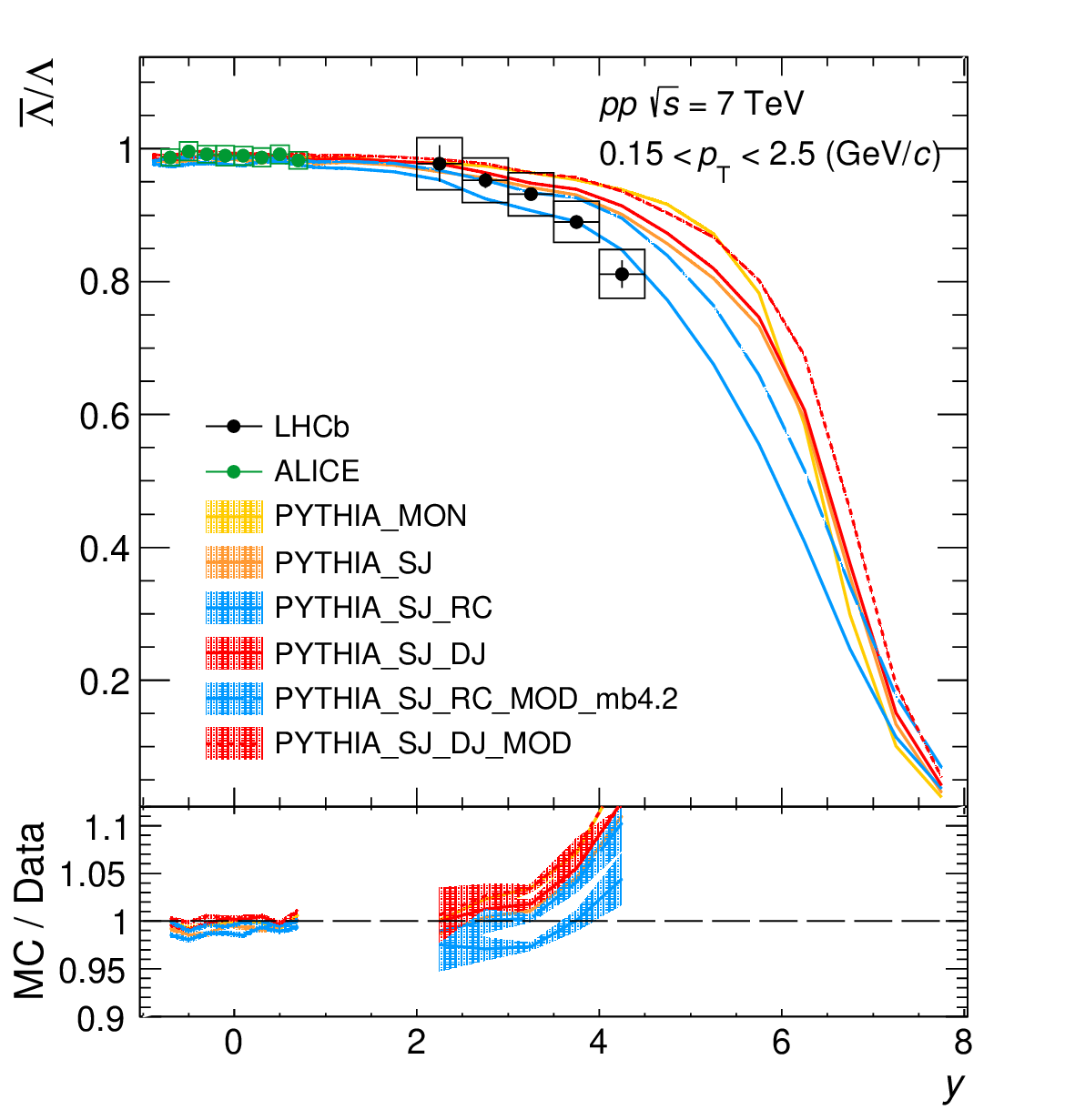}
  \caption{Comparison of rapidity-dependent $\bar{\Lambda}/\Lambda$ ratio measured by ALICE \cite{Abbas2013-dp} and LHCb \cite{Aaij2011-at} in p+p collisions at $\sqrt{s}=$ 7 TeV  with various model predictions. The error bars associated with the data-points represent the statistical uncertainty, whereas the square boxes around the data-points represent the systematic uncertainty}
  \label{fig:aLbyLratio_y_7tev}
\end{figure}

To check the robustness of our {\large P}YTHIA\_SJ\_RC\_MOD\_mb4.2 model generated data, the ALICE and LHCb experimental results on the measurement of $\bar{\Lambda}/\Lambda$ ratio as a function of $y$ for p+p collisions at $\sqrt{s} = $ 7 TeV is compared with our model prediction and is plotted in Fig.\ref{fig:aLbyLratio_y_7tev}. From this figure, it is readily evident that the MC dataset generated with {\large P}YTHIA\_SJ\_RC\_MOD\_mb4.2 shows even a better agreement than the results reported in \cite{PhysRevC.111.014905} with the experimental $\bar{\Lambda} / \Lambda$ ratio at ALICE and LHCb acceptances ($y < 3.5$). The disagreement between the generated and experimental data points is well within 10\% of the uncertainties. \\
Thus, {\large P}YTHIA\_SJ\_RC\_MOD\_mb4.2 model generated dataset has been found to be successful in describing the experimental rapidity-dependent $\bar{\Lambda}/\Lambda$ ratio, $p_\mathrm{T}$-differential production cross-section of $\Lambda_c^+$, $D^0$, $\Lambda_b^0$, $\bar{B^0}$, $p_\mathrm{T}$-dependent ratio of $\Lambda_c^+/D^0$ and $\Lambda_b^0/\bar{B}^0$ within LHCb acceptance in p+p collisions at $\sqrt{s} = $ 7 TeV, as well as, $p_\mathrm{T}$-dependent $\Lambda_b^0/B^0$ ratio of LHCb experiment at  $\sqrt{s} = $ 13 TeV. However, ALICE experiment results on $p_\mathrm{T}$-differential production cross-section of $\Lambda_c^+$, $D^0$ at $\sqrt{s} = $ 7 TeV and $p_\mathrm{T}$-dependent  $\Lambda_c^+/D^0$ at both $\sqrt{s} = $ 7 and 13 TeV were best described by {\large P}YTHIA\_SJ\_DJ\_MOD.

\section{Summary}
In this paper, various MC-generated datasets with different tunings of the PYTHIA 8.314 event generator, based on the string junction, a Y-shaped fundamental topological entity of QCD, are used to study the available experimental $p_\mathrm{T}$-differential production cross sections of heavy-flavored baryons ($\Lambda_c^+$ and $\Lambda_b^0$) and mesons ($D^0$ and $\bar{B}^0 (B^0)$) in p+p collisions at $\sqrt{s} = $ 7 TeV in both ALICE and LHCb acceptances. Moreover, the $p_\mathrm{T}$-dependent ratios of  $\Lambda_c^+/D^0$ and $\Lambda_b^0/\bar{B}^0(B^0)$ at $\sqrt{s} = $ 7 and 13 TeV have also been studied and compared. From the discussions on the charm sector, one can conclude that the dataset {\large P}YTHIA\_SJ\_DJ\_MOD, with fragmentation parameters ($a = 1.1$, $b = 0.6$, $r_c = 1.5$, and $probQQtoQ = 0.12$), is the most successful model in describing the ALICE experimental results on $p_\mathrm{T}$-differential production cross sections of $\Lambda_c^+$ and $D^0$ in p+p collisions at $\sqrt{s} = 7$ TeV, as well as their $p_\mathrm{T}$-dependent ratios at both $\sqrt{s} = 7$ and 13 TeV. However, among all generated datasets, {\large P}YTHIA\_SJ\_RC\_MOD provides the most reasonable description of the $p_\mathrm{T}$-differential production cross sections of $\Lambda_c^+$ and $D^0$, along with the $p_\mathrm{T}$-dependent $\Lambda_c^+/D^0$ ratio in p+p collisions at $\sqrt{s} = $ 7 TeV in the LHCb acceptance. Moreover, the $p_\mathrm{T}$-differential production cross sections of $\Lambda_b^0$, $\bar{B}^0$, and the $p_\mathrm{T}$-dependent $\Lambda_b^0/\bar{B}^0$ ($B^0$) ratio in p+p collisions at $\sqrt{s} = 7$ and 13 TeV in the LHCb acceptance could also be well described by {\large P}YTHIA\_SJ\_RC\_MOD\_mb4.2 data set with a reduced bottom quark mass ($m_b = 4.2 \; \text{GeV}/c^2$) and Bowler's multiplicative factor for bottom quark ($r_b$) set to 0.96. Here, it is worth mentioning that the reduction in the bottom quark mass and modification to the Bowler's multiplicative factor for bottom quark in this dataset have no significant effect on the observables studied in the charm sector. Thus, the model {\large P}YTHIA\_SJ\_DJ\_MOD (by not allowing junction-antijunction reconfiguration) could successfully describe the ALICE experimental observables of the present investigation of the charm sector, whereas the model {\large P}YTHIA\_SJ\_RC\_MOD\_mb4.2 (by allowing junction-antijunction reconfiguration) can explain the considered LHCb experimental observables of both charm and bottom sectors. However, a single set of tuning parameters that can describe all the observables of the present investigation for both ALICE and LHCb results is yet to be realized.

\section{ACKNOWLEDGMENTS}
The authors gratefully acknowledge financial support provided by the Department of Science and Technology, Ministry of Science and Technology, India, under Project No. SR/MF/PS-02/2021-GU(E-37122) to carry out this work.

\end{document}